\documentclass[lettersize,journal]{IEEEtran}
\usepackage{amsmath,amsfonts}
\usepackage{array}
\usepackage[caption=false,font=normalsize,labelfont=sf,textfont=sf]{subfig}
\usepackage{textcomp}
\usepackage{stfloats}
\usepackage{url}
\usepackage{verbatim}
\usepackage{graphicx}
\usepackage{multirow}
\usepackage{booktabs}
\usepackage{algorithm}
\usepackage{amsmath}
\usepackage{algpseudocode}
\usepackage[numbers,sort&compress]{natbib}
\def\BibTeX{{\rm B\kern-.05em{\sc i\kern-.025em b}\kern-.08em
		T\kern-.1667em\lower.7ex\hbox{E}\kern-.125emX}}
\usepackage{balance}
\usepackage{comment}

\begin{document}

\title{Encrypted Semantic Communication Using Adversarial Training for Privacy  Preserving}

\author{Xinlai Luo, Zhiyong Chen, Meixia Tao,~\IEEEmembership{Fellow, IEEE}, and Feng Yang
\thanks{X. Luo, Z. Chen, M. Tao and F. Yang are with the School of Electronic Information and Electrical Engineering, Shanghai Jiao Tong University, China (e-mail: \{newcomer, zhiyongchen, mxtao, yangfeng\}@sjtu.edu.cn).}
}

{


\maketitle

\begin{abstract}
Semantic communication is implemented based on shared background knowledge, but the sharing mechanism risks privacy leakage. In this letter, we propose an encrypted semantic communication system (ESCS) for privacy preserving, which combines universality and confidentiality. The universality is reflected in that all network modules of the proposed ESCS are trained based on a shared database, which is suitable for large-scale deployment in practical scenarios. Meanwhile, the confidentiality is achieved by symmetric encryption. Based on the adversarial training, we design an adversarial encryption training scheme to guarantee the accuracy of semantic communication in both encrypted and unencrypted modes. Experiment results show that the proposed ESCS with the adversarial encryption training scheme can perform well regardless of whether the semantic information is encrypted. It is difficult for the attacker to reconstruct the original semantic information from the eavesdropped message.
\end{abstract}

\begin{IEEEkeywords}
Encrypted semantic communication, symmetric encryption, adversarial training.
\end{IEEEkeywords}

\section{Introduction}
\IEEEPARstart{S}{emantic} communication is built on a common background knowledge base, where both communications nodes privatize the same background knowledge base. The shared background knowledge and the privately trained semantic codec can provide a barrier to privacy protection. Even if a third node eavesdrops on the transmitted semantic message, it is difficult for him to reconstruct the original semantic content based on other background knowledge bases. In this case, semantic communication has high confidentiality but poor generality.
A private communication model must be established between any two communication agents and jointly train a private semantic encoder and decoder. Such a semantic system would be highly complex and challenging to deploy in practical scenarios. 

Therefore, most current research supports a centralized semantic communication system, a unified multi-user semantic communication system trained based on one or several standard background knowledge bases \cite{2020semantic,honggang,2017deep}.
In \cite{2020semantic}, all agents participate in a model update through federated learning to train a generalized semantic model. Through collaborative learning, the model can significantly improve its utility, but it also inevitably suffers from the problem of privacy leakage \cite{2017deep}. Hence, balancing the generality and confidentiality of semantic communication is one of the major challenges of semantic communication.

Recently, neural networks (NNs) based semantic communication model has been proposed. In \cite{semantic}, the authors design an end-to-end (E2E) model for textual semantic communication and apply transfer learning to the training process, significantly reducing the training time. The work \cite{luo} proposes a relay semantic communication model and designs a semantic-and-forward (SF) scheme to solve the heterogeneous background knowledge problem of E2E semantic communication, but it lacks the protection of privacy. In \cite{zhang}, a data adaptation network is proposed to solve the problem of background knowledge heterogeneity and protect the privacy of the pragmatic use of transmitted image data at the receiver. In \cite{learning}, the authors demonstrate that NNs can learn to preserve communication and secure the transmission of the bitstream through adversarial neural encryption.


In this paper, we pay more attention to whether the encoded information can be obtained by eavesdroppers in semantic communication. We propose a universal semantic communication model with a semantic encryption function for the text communication task, termed an encrypted semantic communication system (ESCS). The proposed ESCS provides two modes of semantic transmission, encrypted and unencrypted, without changing the semantic encoder and decoder. Moreover, we design the structure of the secret key, encryptor, and decryptor for semantic communication, where they can be successfully embedded in the shared semantic communication model. Finally, an adversarial encryption training scheme is used to effectively guarantee the accuracy of semantic communication in both encrypted and unencrypted modes and resist attackers from eavesdropping on semantic information \cite{learning,x,cpgan}. Simulation results verify that the proposed ESCS with adversarial training can effectively protect privacy.

\begin{figure*}[t]
	\begin{center}
		\includegraphics[width=17.5cm]{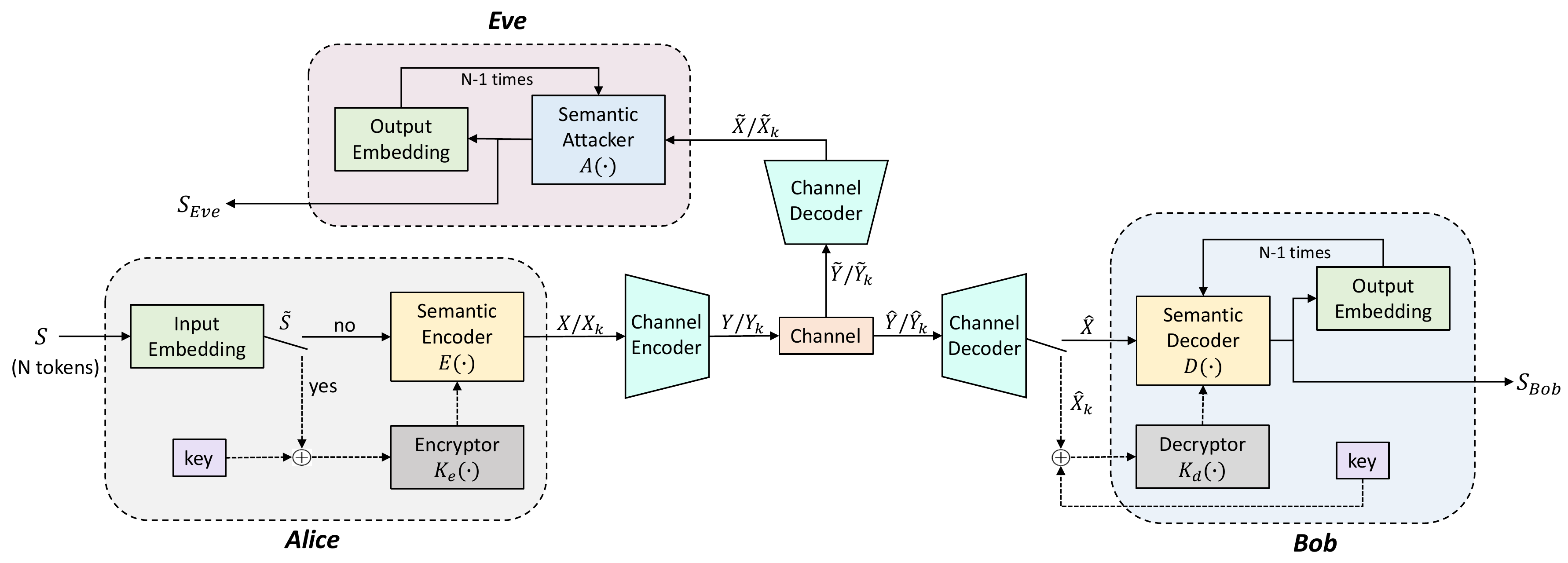}
	\end{center}
	\caption{\small{The structure of the proposed encrypted semantic communication system.}
	}
	\label{model1}
\end{figure*}

\section{System Model}
In this section, we design an encrypted semantic communication system and present how a secret key is used to protect the confidentiality of semantic information.
\subsection{Semantic Symmetric Cryptosystem}
We consider the classic scenario in the security field, which involves three users (Alice, Bob, and Eve). Alice and Bob want to achieve secure semantic communication, and Eve tries to eavesdrop on their communication. The security property mainly prevents eavesdropping because the adversary Eve is limited to intercepting information and cannot inject or modify transmitted messages.

As shown in Fig. \ref{model1}, Alice wishes to send Bob a confidential semantic message $S$. The semantic message $S$ is the input to Alice, and Alice can process the input through the semantic and channel encoder to produce a new message $Y$. Generally, we call this new message ``ciphertext" in the classical encryption scenario. In the semantic communication system, Alice not only encodes the message $S$ semantically but also encrypts the message $S$, so we use $X_{k}$ to represent Alice's encrypted semantic message, where the subscript $k$ represents key encryption. Then, after channel coding, $Y_{k}$ is transmitted over a wireless channel, Bob receives $\hat{Y}_{k}$ and Eve receives $\tilde{Y}_{k}$. They both process the received message and try to recover $S$. We use $S_{Bob}$ and $S_{Eve}$ to represent their recovered results, respectively. Bob has one advantage over Eve in that he shares a secret key with Alice. We treat the secret key as additional input for Alice and Bob. Each semantic message $S$ matches a new key during communication.
\subsection{Encrypted Semantic Communication System}
We also consider both universality and confidentiality of the proposed encrypted semantic communication system. Thus, the semantic encryption function of this system is optional. If the message does not require privacy protection, the message can be transmitted in an unencrypted way. Since the semantic encoding module is generic, anyone can decode it in this case. This universality of the system is suitable for various scenarios, such as broadcast channels. Moreover, the training of the system model is unified, which improves the efficiency of model training and reduces the deployment complexity of semantic communication components.

The proposed system requires different semantic coding networks for different input forms. This paper mainly introduces text-type input. The input is a sentence, and we tokenize the input sentence as $S$. Each token in $S$ is a one-hot vector whose length is the size of the word dictionary in the background knowledge base. Through the word embedding layer, we can map each token to a fixed-dimensional vector of floats, and the output is $\tilde{S}$. Then, the transmitter can choose whether to encrypt the semantic message. If it needs to encrypt,  input $\tilde{S}$ together with the key to the encryptor $K_{e}(\cdot)$ for encryption, and then input the encrypted message to the semantic encoder $E(\cdot)$ for semantic encoding. Otherwise, input $\tilde{S}$ directly to the semantic encoder $E(\cdot)$ for semantic encoding. The semantic encoder encodes the semantic message, whether encrypted or not and then outputs the semantic vector $X$ and $ X_{k} $ respectively. Finally, the system performs the channel coding on the semantic vector $X$ or $X_{k}$ to obtain the output  $Y$ or $Y_{k}$.
In this paper, we use the transformer network as the semantic codec \cite{attention}, which performs well on text tasks. Autoencoder can be used as the channel codec \cite{luo}. 

The network structure of the encryptor and decryptor is the same, as shown in Fig. \ref{model2}. The original semantic message has $N$ tokens, and each token is embedded into $M$ dimensions. The secret key is a random vector of $M$-dimensional floating-point numbers, equivalent to a token in the semantic message. Both of them are input to the encryptor or decryptor. The first layer of the encryptor is the reshape layer, which concatenates the original message and the key and then does a dimensional transformation. The second layer is the hidden layer, and the dimension is reduced to the length of the original message. Finally, the output reshapes the vector to be the same as the original message. It is worth noting that the form of encryption and decryption is discovered by learning, not generated by a specific algorithm. The reason why the key chooses a token length is that the length of the key does not represent the strength of the disturbance. The effect of encryption lies in the encryption method learned by the network, not the amount of data in the key. In addition, the location choices of the encryptor and decryptor are also various, which can make reasonable adjustments, but the network performance is not much different. We have also verified these two points through experiments.

In this paper, we consider the additive white Gaussian noise (AWGN) channel, where one output of the channel is denoted as $\hat{Y} = Y + {N}_{t}$, where $N_{t} \sim \mathcal{ N}\left(0, \sigma^{2}\right)$. Based on the received signal $\tilde{Y}$ or $\tilde{Y}_{k}$, Eve tries to reconstruct the original semantic message by semantic attacker $A(\cdot)$. Meanwhile, Bob decodes the unencrypted message directly through the semantic decoder $D(\cdot)$, while the encrypted message is first decrypted with the decryptor $K_{d}(\cdot)$, and then decoded through the semantic decoder $D(\cdot)$. Note that since the attacker may be the receiver in other communication links, we consider the receiver and the attacker have the same semantic decoder structure. The semantic decoder decodes word by word, so the output of the first $N-1$ times can be used as another input to the semantic decoder. In other words, the number of decoding operations the decoder will perform is equal to the number of tokens in the original input.

\begin{figure}[t]
	\begin{center}
		\includegraphics[width=8.5cm]{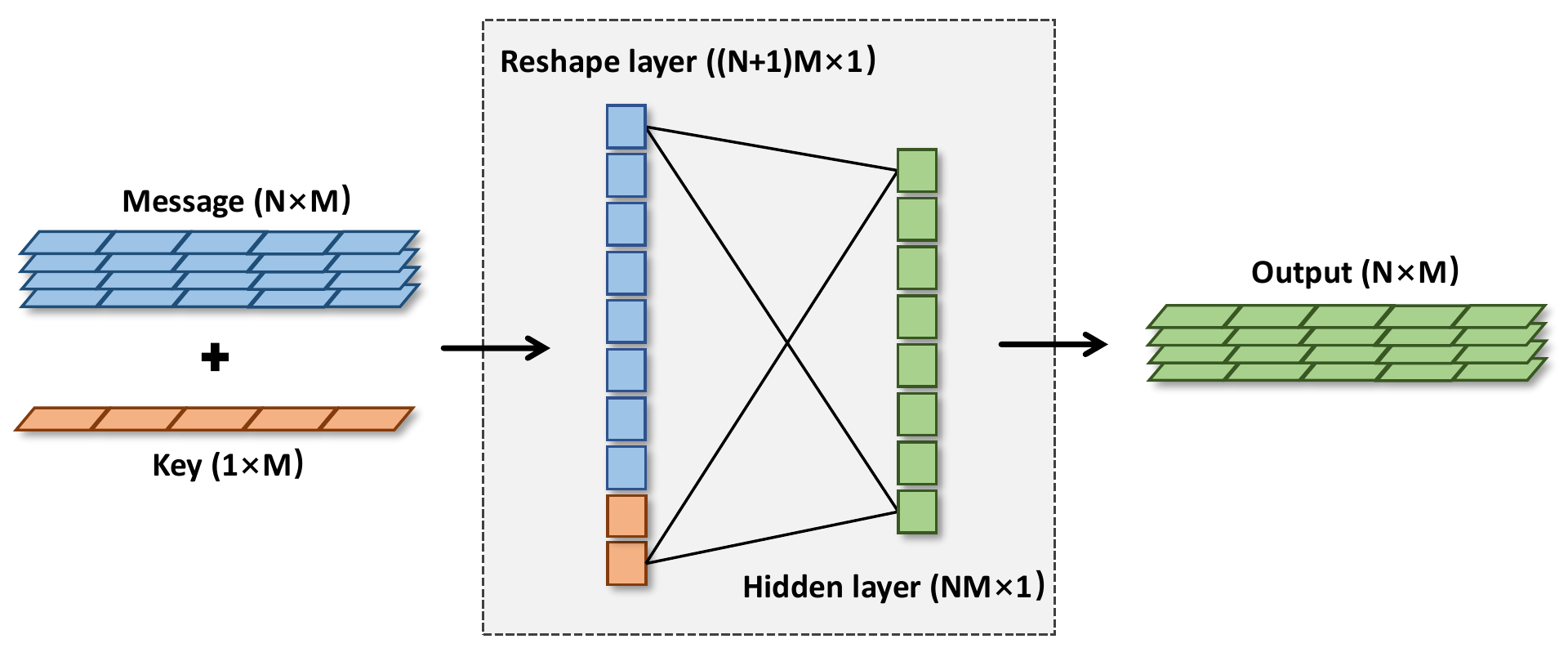}
	\end{center}
	\caption{\small{Encryptor and decryptor network structure.}
	}
	\label{model2}
\end{figure}

\section{Objectives and Training}
In this section, we describe the objectives of each participant in the ESCS in detail, design loss functions for different network modules based on the objectives, and use specific training methods to achieve optimal performance.

\subsection{Loss Function Design}
If the content of the communication is not confidential, the attacker Eve is ignored so that the objective is to minimize the error between $S$ and $S_{Bob}$. If Alice and Bob want to hide the communication content from Eve, the goal is to minimize the error between $S$ and $S_{Bob}$ while maximizing the error between $S$ and $S_{Eve}$. For Eve, the goal is to accurately reconstruct $S$, that is, to minimize the error between $S$ and $S_{Eve}$. As a result, we defeat the attacker by jointly training the transmitter and receiver, where the eavesdropping ability of the attacker is also enhanced during training. Similar to generative adversarial networks (GANs) \cite{GAN}, we want the transmitter and receiver to beat the best attackers, not a fixed one.

The following description only presents the network parameters that need to be updated in the loss function. The embedding layer is fixed and the channel encoding and decoding networks are pre-trained \cite{luo}. For the distance function, we use cross-entropy $\mathcal{D}_{C E}$, which can be formulated as
\begin{equation}
\mathcal{D}_{C E}=-\frac{1}{m} \sum_{i=1}^{m} \sum_{j=1}^{l} p\left(x_{i j}\right) \log \left(q\left(x_{i j}\right)\right),
\end{equation}
where $p\left(x_{i j}\right)$ and $q\left(x_{i j}\right)$ are the real probability and the predicted probability, respectively, of the $j$-th word in yhe $i$-th sample, $l$ represents the number of tokens in the sentence, and $m$ represents the number of samples in one batch.

Firstly, the loss function for unencrypted semantic communication is given by
\begin{equation}
	\label{1}
	\mathcal{L}_{ED}\left(\theta_{E}, \theta_{D}\right)=\mathcal{D}_{C E}\left(S, D\left(\theta_{D}, E\left(\theta_{E}, S\right)\right)\right),
\end{equation}
where $E(\theta_{E},\cdot)$ and $D(\theta_{D},\cdot)$ represent the outputs of the semantic encoder and semantic decoder, respectively. Note that $E\left(\theta_{E}, S\right)$ is not directly the input of $ D(\cdot)$ during training, it needs to pass through the channel, but in order to simplify the expression, it is not expressed in the formula. Therefore, we have the optimal semantic encoder and decoder by minimizing this loss as following
\begin{equation}
	\label{2}
	\left(O_{E}, O_{D}\right)=\operatorname{argmin}_{\left(\theta_{E}, \theta_{D}\right)}\left(\mathcal{L}_{ED}\left(\theta_{E}, \theta_{D}\right)\right).
\end{equation}
Similarly, we define a loss function for the encrypted semantic communication as following
\begin{equation}
	\label{3}
	\begin{aligned}
		\mathcal{L}_{K_{d}}\left(\theta_{K_{e}}, \theta_{E}, \theta_{K_{d}}, \theta_{D}\right)=& \mathcal{D}_{C E}\left(S, D\left(\theta_{D}, K_{d}\left(\theta_{K_{d}},\right.\right.\right.\\
		& E\left(\theta_{E}, K_{e}\left(\theta_{K_{e}}, S\right)\right)))),
	\end{aligned}
\end{equation}
where $K_{e}(\theta_{K_{e}},\cdot)$ and $K_{d}(\theta_{K_{d}},\cdot)$ are the outputs of the encryptor and decryptor, respectively. We obtain the optimal decryptor for receiver by minimizing the loss as following
\begin{equation}
	\label{4}
	O_{K_{d}}\left(\theta_{K_{d}}\right)=\operatorname{argmin}_{\theta_{K_{d}} }\left(\mathcal{L}_{K_{d}}\left(\theta_{K_{e}}, \theta_{E}, \theta_{K_{d}}, \theta_{D}\right)\right).
\end{equation}

The attacker would intercept the encrypted message and reconstruct the semantic information directly using a semantic attacker. The loss function for the attacker can be given by
\begin{equation}
	\label{5}
	\mathcal{L}_{A}\left(\theta_{K_{e}}, \theta_{E}, \theta_{A}\right)=\mathcal{D}_{C E}\left(S, A\left(\theta_{A}, E\left(\theta_{E}, K_{e}\left(\theta_{K_{e}}, S\right)\right)\right)\right),
\end{equation}
where $A(\theta_{A},\cdot)$ is the output of the semantic attacker. The optimal attacker can be obtained by minimizing the loss as following
\begin{equation}
	\label{6}
	O_{A}\left(\theta_{A}\right)=\operatorname{argmin}_{\theta_{A} }\left(\mathcal{L}_{A}\left(\theta_{K_{e}} \theta_{E}, \theta_{A}\right)\right).
\end{equation}
Therefore, a loss function for encryptor and decryptor by combining $L_{K_{d}}$ and $L_{A}$ can be given by
\begin{equation}
	\label{7}
	\begin{aligned}
	\mathcal{L}_{K_{e}}\left(\theta_{K_{e}}\right)= &\mathcal{L}_{K_{d}}\left(\theta_{K_{e}} \theta_{E}, \theta_{K_{d}}, \theta_{D}\right)\\
	&-\lambda \mathcal{L}_{A}\left(\theta_{K_{e}}, \theta_{E}, O_{A}\left(\theta_{A}\right)\right).
    \end{aligned}
\end{equation}
Here, the hyper-parameter $\lambda$ balances the utility and confidentiality. We obtain the optimal encryptor by minimizing this loss
\begin{equation}
	\label{8}
	O_{K_{e}}\left(\theta_{K_{e}}\right)=\operatorname{argmin}_{\theta_{K_{e}} }\left(\mathcal{L}_{K_{e}}\left(\theta_{K_{e}} \right)\right).
\end{equation}

Generally speaking, the transmitter and receiver can have many near-optimal solutions because the encryption and decryption methods are not fixed during the learning process, the size of the key is variable, and the value of the key is random. We will explain the detailed training process in the following subsection.

\subsection{Training Refinement}
\begin{algorithm}[t]
	\caption{ESCS training algorithm} 
	\hspace*{0.02in} {\bf \small{Input:}} 
    \small{	Channel SNR value and hyper-parameter $\lambda$.} \\
	\hspace*{0.02in} {\bf \small{Output:}} 
	\small{Network $K_{e}(\mathbf{\cdot})$, $K_{d}(\mathbf{\cdot})$, $E(\mathbf{\cdot})$, $D(\mathbf{\cdot})$, $A(\mathbf{\cdot})$.}
	\label{whole}
	\begin{algorithmic}[1]
		\small{\State Load the pre-trained channel encoder and decoder.
		\State Load and fix the parameters of the embedding network.
		\State Set epoch counter $t = 1$.
		\While{the training stop condition is not met}
		\State Take a batch $S$ from the set and embed it as $\tilde{S}$;
		\State Randomly generate a key ;
		\State Semantic encryption encode $ X_{k} = E(K_{e}(\tilde{S}))$ by key;
		\State Semantic encode  $ X = E(\tilde{S})$;
		\State Transmit $Y/Y_{k}$ over the channel;
		\State Decode $D(\hat{X})$ to compute loss $\mathcal{L}_{ED}$ (receiver);
		\State Decode $D(K_{d}(\hat{X}_{k}))$ to compute loss $\mathcal{L}_{K_{d}}$ (receiver);
		\State Decode $A(\tilde{X}_{k})$ to compute loss $\mathcal{L}_{A}$ (attacker);
		\State Compute loss $\mathcal{L}_{K_{e}} = \mathcal{L}_{K_{d}} - \lambda \mathcal{L}_{A}$ (transmitter).
		\If{$t \mod  4 = 0 $} 
		\State Gradient descent update $\theta_{E}, \theta_{D}$ to minimize $\mathcal{L}_{ED}$.
		\ElsIf{$t \mod  4 = 1 $} 
		\State Gradient descent update $\theta_{K_{d}}$ to minimize $\mathcal{L}_{K_{d}}$.
		\ElsIf{$t \mod  4 = 2 $}
		\State Gradient descent update $\theta_{A}$ to minimize $\mathcal{L}_{A}$.
		\Else
		\State Gradient descent update $\theta_{K_{e}}$ to minimize $\mathcal{L}_{K_{e}}$.
		\EndIf
		\State $t = t + 1$.
		\EndWhile}
	\end{algorithmic}
\end{algorithm}
The training of the encrypted semantic communication network is divided into two steps. The first is to train the channel encoder and decoder with a symmetric structure. They each have two hidden layers, each of which compresses the input vector to a certain extent and finally maps it to the symbol with a real part and an imaginary part. We use randomly generated vectors for training, similar to the encoded semantic vector. The channel parameters are set to dynamically change within a certain range during training, which can enhance the robustness. And the mean square error (MSE) is used as the loss function to reduce the distortion.

Similar to GANs, we alternately train the attacker with transmitter and receiver. Intuitively, the training algorithm is roughly outlined in Algorithm \ref{whole}. With a few steps of training, the semantic encoder $E(\mathbf{\cdot})$ and decoder $D(\mathbf{\cdot})$ find a way to satisfy common semantic communication requirements. Furthermore, the decryptor $K_{d}(\mathbf{\cdot})$ continues to learn a stable decryption method, but the semantic attacker $A(\mathbf{\cdot})$ gradually learns a way to decode the encrypted semantic message directly. In this process, the receiver and the attacker attempt to minimize reconstruction error at the same time, so we alternately update the encryptor $K_{e}(\mathbf{\cdot})$ to reduce the receiver's reconstruction error but increase the attacker's reconstruction error. Intuitively, learning makes the encryption method more receiver friendly.

\section{Performance Evaluation}
In this section, we present numerical results to evaluate the performance of the proposed ESCS scheme. The dataset in the experiments is the standard proceedings of the European Parliament \cite{corpus}, which consists of around 2.0 million sentences. The learning rate for Algorithm 1 is set to $10^{-4}$, hyper-parameter $\lambda$ is set to $0.2$, and the signal-to-noise ratio(SNR) of the wireless channel is set to 10 dB. We apply the bilingual evaluation understudy (BLEU) score \cite{bleu} as the evaluation metrics. The weights of 1-gram and 2-gram in BLEU are set to 0.6 and 0.4, respectively. In the experiment, we compare the accuracy of the reconstructed semantic messages between the receiver and the attacker based on the proposed ESCS scheme and the non-adversarial encryption training scheme.

Fig. \ref{f3} shows the change of $\mathcal{L}_{ED}$, $ \mathcal{L}_{K_{d}} $ and $\mathcal{L}_{A}$ with the increase of training steps using adversarial encryption training or non-adversarial encryption training. We can see that with the proposed ESCS scheme, $\mathcal{L}_{ED}$ and $ \mathcal{L}_{K_{d}}$ can converge to 0. However, $\mathcal{L}_{A}$ cannot converge to 0 and eventually fluctuate around 0.5. Because the cross-entropy is used as the distance function, if the loss value cannot converge to 0, the decoder will not be able to reconstruct the original message correctly. On the other hand, using the non-adversarial training scheme, that is, when $\mathcal{L}_{K_{e}}$ does not participate in the training update, $\mathcal{L}_{ED}$, $ \mathcal{L}_{K_{d}} $ and $\mathcal{L}_{A}$ can all converge to 0. It means that although the system has high utility, confidentiality is inferior. Eve can easily reconstruct the original semantic message from the encrypted messages.

We can also see from Fig. \ref{f3} that with the adversarial encryption training scheme, the convergence speed of each loss value is slower than that of the non-adversarial encryption training scheme. To defeat the best attackers, the encryptor will update its network parameters after a certain number of training steps. This update is suitable for Bob but not friendly to Eve, so Eve's loss value fluctuates wildly. More importantly, such an adversarial training process makes all loss values fall more slowly and eventually prevents Eve's loss from converging to 0. 

\begin{figure*}[t]
	\centering
	\subfloat[\small{$\mathcal{L}_{ED}$, $ \mathcal{L}_{K_{d}} $ and $\mathcal{L}_{A}$ from left to right} \label{fig:a}]{
		\includegraphics[width=14.5cm]{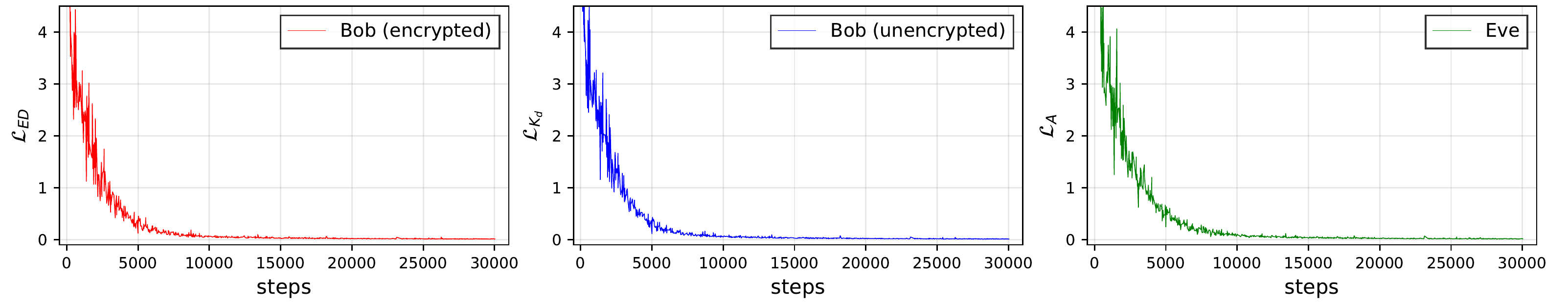}}\\
	\subfloat[\small{$\mathcal{L}_{ED}$, $ \mathcal{L}_{K_{d}} $ and $\mathcal{L}_{A}$ from left to right}\label{fig:b}]{
		\includegraphics[width=14.5cm]{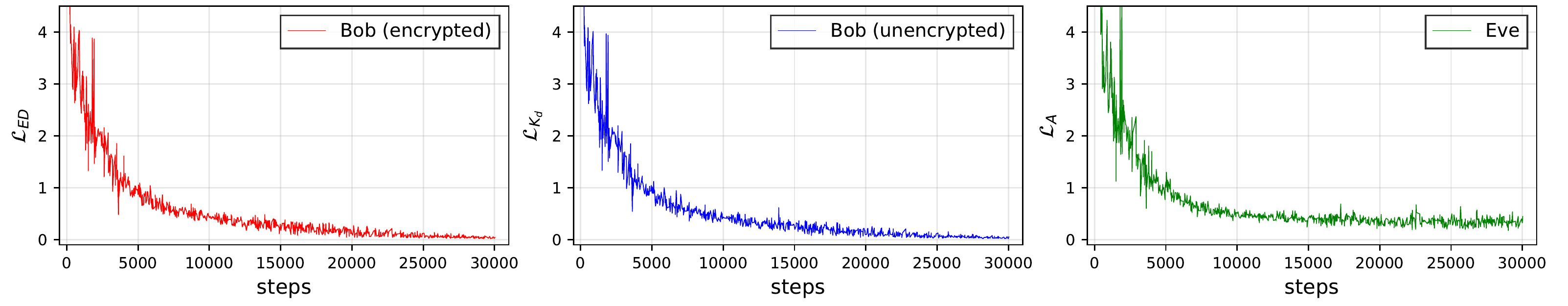}}
	\caption{\small{Training scheme of ESCS: (a) Adversarial encryption training. (b) Non-adversarial encryption training.}
	}
	\label{f3}
\end{figure*}


\begin{figure}[t]
	\begin{center}
		\includegraphics[width=6.5cm]{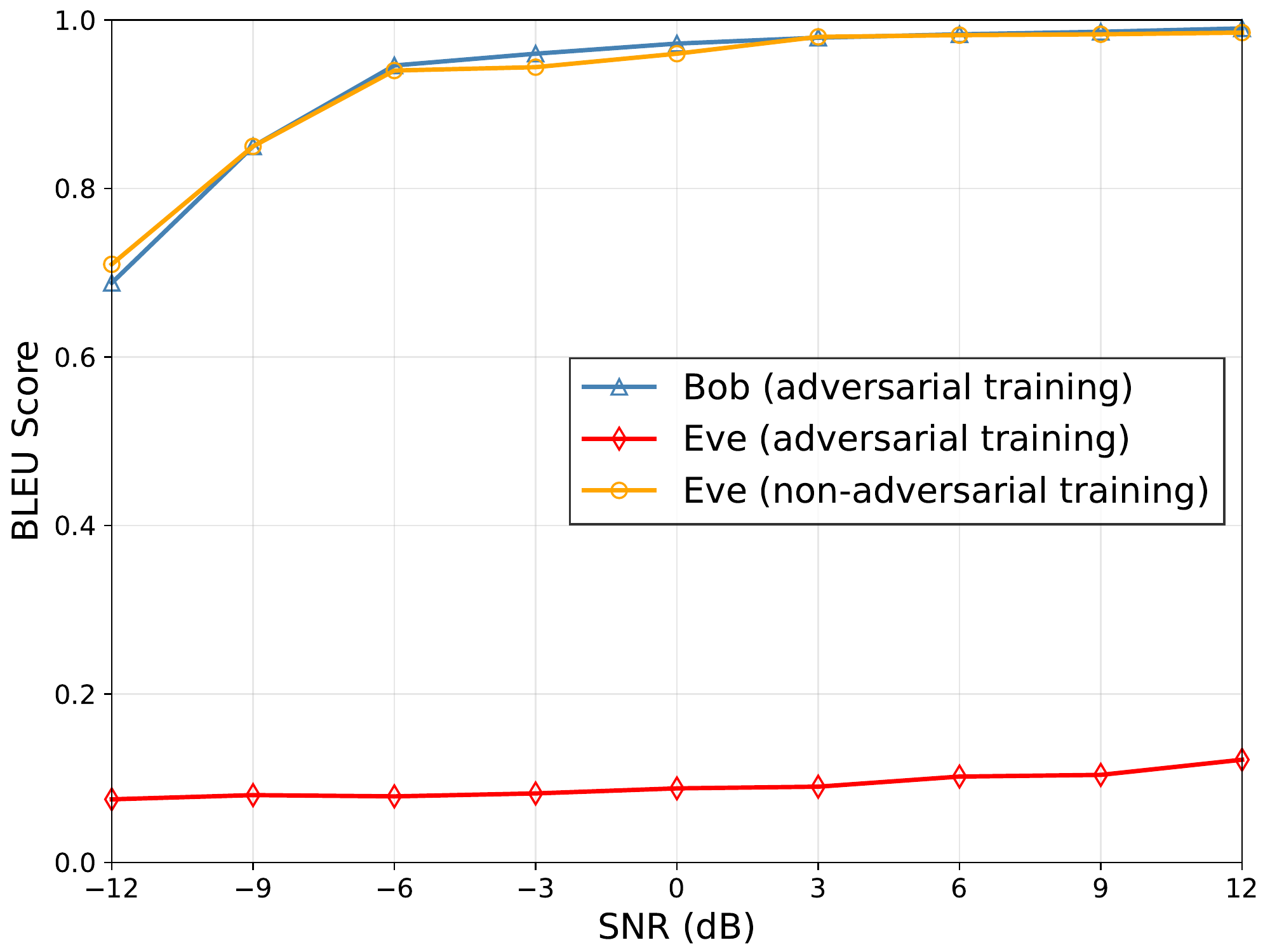}
	\end{center}
	\caption{\small{BLEU score versus SNR for receiver (Bob) and attacker (Eve) in ESCS with different training schemes.}
	}
	\label{f4}
\end{figure}

In Fig. \ref{f4}, we show the BLEU score vs. SNR for different training schemes. We can see that the BLEU score of Bob is much higher than Eve's at any SNR when using the adversarial encryption training scheme. Under high SNR channel conditions, the BLEU score of Bob is close to 1, while the BLEU score of Eve is less than 0.2. As a comparison, we verified the performance of Eve eavesdropping semantic information when training with non-adversarial encryption. At this time, the BLEU score of Eve is almost the same as the BLEU score of Bob, which means that the privacy leakage is severe and proves that adversarial encryption training can effectively protect privacy. 

Fig. \ref{f5} presents the effect of keys of different lengths on ESCS. We can see that the long key can reduce communication performance when the channel condition is poor. Because the encryption method of long keys is complicated, the anti-noise ability of encrypted information is insufficient. So we can use the key with one token to ensure low encryption complexity and the best performance of ESCS\label{key}.

\section{Conclusion}
In this letter, we have applied symmetric encryption to solve the security problem of eavesdropping in the semantic communication system. To make the proposed ESCS both universal and confidential, we have proposed an adversarial encryption training scheme, which can effectively guarantee the accuracy of semantic communication in both encrypted and unencrypted modes and resist attackers from eavesdropping on semantic information. Simulation results have demonstrated that the proposed ESCS using the adversarial training scheme can significantly improve the privacy protection capability of the semantic communication system.

\begin{figure}[t]
	\begin{center}
		\includegraphics[width=6.5cm]{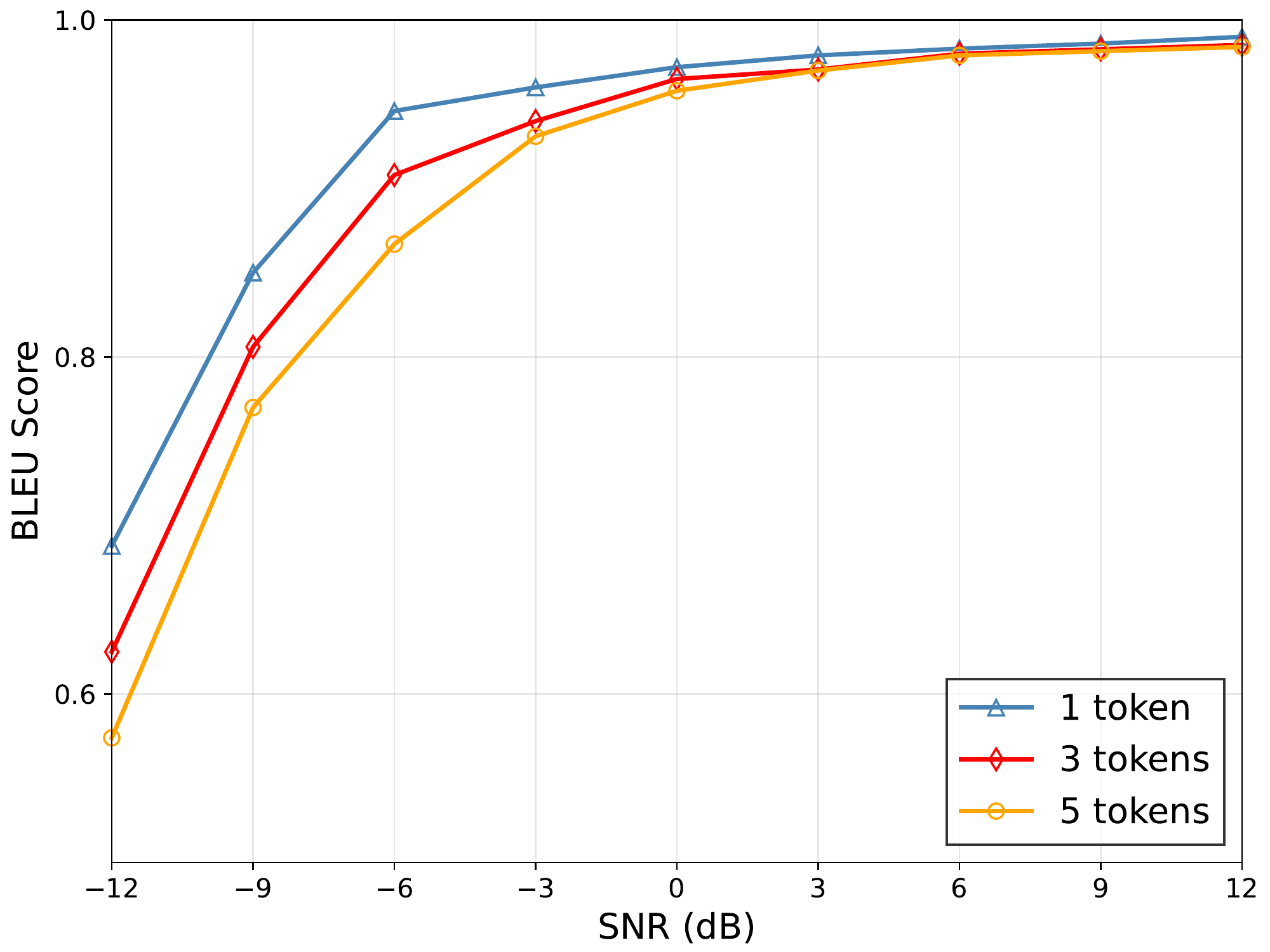}
	\end{center}
	\caption{\small{BLEU score versus SNR for receiver (Bob) when the key takes different lengths. }
	}
	\label{f5}
\end{figure}
\small
\bibliographystyle{IEEEtran}
\bibliography{ref}{}

\begin{thebibliography}{10}
\providecommand{\url}[1]{#1}
\csname url@samestyle\endcsname
\providecommand{\newblock}{\relax}
\providecommand{\bibinfo}[2]{#2}
\providecommand{\BIBentrySTDinterwordspacing}{\spaceskip=0pt\relax}
\providecommand{\BIBentryALTinterwordstretchfactor}{4}
\providecommand{\BIBentryALTinterwordspacing}{\spaceskip=\fontdimen2\font plus
\BIBentryALTinterwordstretchfactor\fontdimen3\font minus
  \fontdimen4\font\relax}
\providecommand{\BIBforeignlanguage}[2]{{%
\expandafter\ifx\csname l@#1\endcsname\relax
\typeout{** WARNING: IEEEtran.bst: No hyphenation pattern has been}%
\typeout{** loaded for the language `#1'. Using the pattern for}%
\typeout{** the default language instead.}%
\else
\language=\csname l@#1\endcsname
\fi
#2}}
\providecommand{\BIBdecl}{\relax}
\BIBdecl

\bibitem{2020semantic}
G.~Shi, Y.~Xiao, Y.~Li, and X.~Xie, ``From semantic communication to
  semantic-aware networking: Model, architecture, and open problems,''
  \emph{IEEE Communications Magazine}, vol.~59, no.~8, pp. 44--50, 2021.

\bibitem{honggang}
Q.~Zhou, R.~Li, Z.~Zhao, C.~Peng, and H.~Zhang, ``Semantic communication with
  adaptive universal transformer,'' \emph{IEEE Wireless Communications
  Letters}, vol.~11, no.~3, pp. 453--457, 2022.

\bibitem{2017deep}
B.~Hitaj, G.~Ateniese, and F.~Perez-Cruz, ``Deep models under the gan:
  information leakage from collaborative deep learning,'' in \emph{Proceedings
  of the 2017 ACM SIGSAC conference on computer and communications security},
  2017, pp. 603--618.

\bibitem{semantic}
H.~Xie, Z.~Qin, G.~Y. Li, and B.-H. Juang, ``Deep learning enabled semantic
  communication systems,'' \emph{IEEE Trans. on Signal Processing}, vol.~69,
  pp. 2663--2675, 2021.

\bibitem{luo}
X.~Luo, Z.~Chen, B.~Xia, and J.~Wang, ``Autoencoder-based semantic
  communication systems with relay channels,'' in \emph{Proc. 2022 IEEE
  International Conference on Communications Workshops (ICC Workshops)}, 2022,
  pp. 1--6.

\bibitem{zhang}
H.~Zhang, S.~Shao, M.~Tao, X.~Bi, and K.~B. Letaief, ``Deep learning-enabled
  semantic communication systems with task-unaware transmitter and dynamic
  data,'' \emph{arXiv preprint arXiv:2205.00271}, 2022.

\bibitem{learning}
M.~Abadi and D.~G. Andersen, ``Learning to protect communications with
  adversarial neural cryptography,'' \emph{arXiv preprint arXiv:1610.06918},
  2016.

\bibitem{x}
J.~M. Perero-Codosero, F.~M. Espinoza-Cuadros, and L.~A.
  Hern{\'a}ndez-G{\'o}mez, ``X-vector anonymization using autoencoders and
  adversarial training for preserving speech privacy,'' \emph{Computer Speech
  \& Language}, p. 101351, 2022.

\bibitem{cpgan}
B.-W. Tseng and P.-Y. Wu, ``Compressive privacy generative adversarial
  network,'' \emph{IEEE Trans. on Information Forensics and Security}, vol.~15,
  pp. 2499--2513, 2020.

\bibitem{attention}
A.~Vaswani, N.~Shazeer, N.~Parmar, J.~Uszkoreit, L.~Jones, A.~N. Gomez,
  {\L}.~Kaiser, and I.~Polosukhin, ``Attention is all you need,''
  \emph{Advances Neural Info. Process. Systems}, vol.~30, 2017.

\bibitem{GAN}
I.~Goodfellow, J.~Pouget-Abadie, M.~Mirza, B.~Xu, D.~Warde-Farley, S.~Ozair,
  A.~Courville, and Y.~Bengio, ``Generative adversarial nets,'' \emph{Advances
  in neural information processing systems}, vol.~27, 2014.

\bibitem{corpus}
P.~Koehn \emph{et~al.}, ``Europarl: A parallel corpus for statistical machine
  translation,'' in \emph{MT summit}, vol.~5.\hskip 1em plus 0.5em minus
  0.4em\relax Citeseer, 2005, pp. 79--86.

\bibitem{bleu}
K.~Papineni, S.~Roukos, T.~Ward, and W.-J. Zhu, ``Bleu: a method for automatic
  evaluation of machine translation,'' in \emph{Proceedings of the 40th annual
  meeting of the Association for Computational Linguistics}, 2002, pp.
  311--318.

\end{thebibliography}
\end{document}